\documentclass{article}
    \usepackage[nonatbib, preprint]{tackling_climate_workshop_style}

\usepackage{comment}
\usepackage[utf8]{inputenc}
\usepackage[T1]{fontenc}
\usepackage{hyperref}
\usepackage{svg}
\usepackage{url}
\usepackage{booktabs}
\usepackage{amsfonts}
\usepackage{amsmath}
\usepackage{nicefrac}
\usepackage{microtype}
\usepackage{cite}
\usepackage{float}
\usepackage{graphicx}
\usepackage{subfig}

\title{\textit{AQ-PINNs: Attention-Enhanced Quantum Physics-Informed Neural Networks for Carbon-Efficient Climate Modeling}}
\author{Siddhant Dutta\textsuperscript{1}, Nouhaila Innan\textsuperscript{2,3}, Sadok Ben Yahia\textsuperscript{4}, Muhammad Shafique\textsuperscript{2,3} \\
\textsuperscript{1}SVKM's Dwarkadas J. Sanghvi College of Engineering, India\\ 
\textsuperscript{2}eBRAIN Lab, Division of Engineering, \\New York University Abu Dhabi (NYUAD), Abu Dhabi, UAE\\
\textsuperscript{3}Center for Quantum and Topological Systems (CQTS), \\NYUAD Research Institute, NYUAD, Abu Dhabi, UAE\\
\textsuperscript{4}The Maersk Mc-Kinney Moller Institute, \\University of Southern Denmark, 
Alsion 2, 6400- Sønderborg, Denmark \vspace{0.2cm}
 \\
\small siddhant.dutta180@svkmmumbai.onmicrosoft.com, nouhaila.innan@nyu.edu, \\ \small say@mmmi.sdu.dk,
muhammad.shafique@nyu.edu}
\begin{document}

\maketitle

\begin{abstract}
The growing computational demands of artificial intelligence (AI) in addressing climate change raise significant concerns about inefficiencies and environmental impact, as highlighted by the Jevons paradox. We propose an attention-enhanced quantum physics-informed neural networks model (AQ-PINNs) to tackle these challenges. This approach integrates quantum computing techniques into physics-informed neural networks (PINNs) for climate modeling, aiming to enhance predictive accuracy in fluid dynamics governed by the Navier-Stokes equations while reducing the computational burden and carbon footprint. By harnessing variational quantum multi-head self-attention mechanisms, our AQ-PINNs achieve a 51.51\% reduction in model parameters compared to classical multi-head self-attention methods while maintaining comparable convergence and loss. It also employs quantum tensor networks to enhance representational capacity, which can lead to more efficient gradient computations and reduced susceptibility to barren plateaus. Our AQ-PINNs represent a crucial step towards more sustainable and effective climate modeling solutions.
\end{abstract}

\section{Introduction}

The rapid growth of artificial intelligence (AI) has led to a significant increase in energy consumption and carbon emissions, primarily due to the escalating computational requirements of increasingly complex and accurate models. This trend, often called the \textit{AI arms race}, has seen energy consumption doubling approximately every 3.4 months since 2012 \cite{lohn2022ai}. For instance, the training of GPT-3, with its 175 billion parameters, has been estimated to emit over an amount of electricity equivalent to 500 metric tons of $CO_2$, a figure comparable to the annual emissions of approximately 60 passenger vehicles \cite{patterson2021carbon}. In response to this unsustainable trajectory, the research community has begun focusing on \textit{Green AI} methodologies and tools like CodeCarbon \cite{courty2024mlco2}, Carbontracker \cite{anthony2020carbontracker}, and eco2AI to quantify and mitigate the environmental impact of AI models throughout their lifecycle \cite{budennyy2022eco2ai}. However, the Jevons paradox presents a crucial challenge: improvements in energy efficiency \cite{Jevons}, such as those offered by Green AI, can paradoxically lead to an overall increase in energy consumption. This is because greater efficiency often leads to broader adoption and use of AI technologies, potentially exacerbating the very environmental impact they aim to mitigate.

Quantum machine learning (QML) benefits from quantum tensor networks by significantly reducing the number of model parameters and leveraging the mathematical essence of Hilbert space to achieve more compact representations \cite{schuld2015introduction,biamonte2017quantum,zaman2023survey}. Quantum tensor networks, such as matrix product states (QMPS), tree tensor networks (QTTN), and multiscale entanglement renormalization ansatz (QMERA) \cite{Huggins_2019}, offer various approaches to represent complex quantum states efficiently. For example, QMPS and QTTN can encode quantum states with fewer parameters compared to fully connected quantum circuits, thereby reducing the model complexity while maintaining high accuracy compared to their classical counterparts. In a specific case involving the MNIST dataset, quantum tensor networks trained with fewer parameters achieved high accuracy compared to a classical neural network \cite{kong2021quantum,liu2024training}. These tensor network architectures enable more efficient data processing and parameterization by exploiting specific entanglement patterns and reducing redundant information. 

In this work, addressing the significant environmental implications, we emphasize the urgent need to develop AI models that achieve high performance while minimizing energy consumption. We propose a novel architecture called \textit{attention-enhanced quantum physics-informed neural networks} (AQ-PINNs), which integrates quantum computing principles with energy-efficient AI frameworks. By employing quantum multi-head self-attention mechanisms alongside quantum tensor networks, AQ-PINNs aim to reduce the parameter space, thus decreasing the computational resources required for training and inference compared to classical models. Our goal is to advance climate modeling capabilities and contribute to sustainable climate change projections, all while maintaining high model performance.

\section{Methodology}

\subsection{Data}

Computational fluid dynamics (CFD) is a vital tool for understanding and predicting fluid flow in various applications, including climate modeling \cite{zawawi2018review}. The dataset used in this study is derived from numerical solutions of the incompressible Navier-Stokes equations \cite{CODINA1999467}. The data is stored in the .mat file and contains the following components:
\begin{itemize}
    \item $X_{star} \in \mathbb{R}^{5000 \times 2}$: Spatial coordinates $(x, y)$ for 5000 points, representing a grid similar to those used in climate models.
    \item $U_{star} \in \mathbb{R}^{5000 \times 2 \times 200}$: Velocity field $(u, v)$, crucial for modeling atmospheric and oceanic currents in climate systems.
    \item $t \in \mathbb{R}^{200 \times 1}$: Time points ranging from 0 to 19.9 seconds, scalable to represent longer climate timescales.
    \item $p_{star} \in \mathbb{R}^{5000 \times 200}$: Pressure field, essential for understanding weather patterns and climate dynamics.
\end{itemize}

The raw data is flattened and reorganized from its original 3D structure into 1D arrays, enabling efficient sampling and manipulation. The training set is created by randomly selecting a subset of the flattened data, ensuring a diverse and representative sample of the entire dataset. Specifically, $``N\_train = 30000"$ training points are chosen without replacement. The entire grid of spatial coordinates is utilized as a test set, with time set to a uniform value for visualization consistency. The incompressible Navier-Stokes equations governing this system key to climate modeling are:

\(
    \frac{\partial \mathbf{u}}{\partial t} + (\mathbf{u} \cdot \nabla)\mathbf{u} = -\frac{1}{\rho}\nabla p + \nu \nabla^2\mathbf{u}, \quad with \quad \nabla \cdot \mathbf{u} = 0,
 \)
where $\mathbf{u} = (u, v)$ is the velocity field (representing wind or ocean currents in climate contexts), $p$ is the pressure, $\rho$ is the fluid density, and $\nu$ is the kinematic viscosity. In our two-dimensional setting, analogous to simplified climate models \cite{jacques2022simplified}, these equations expand to:
\begin{equation}
\small
    \frac{\partial u}{\partial t} + u\frac{\partial u}{\partial x} + v\frac{\partial u}{\partial y} = -\frac{1}{\rho}\frac{\partial p}{\partial x} + \nu\left(\frac{\partial^2 u}{\partial x^2} + \frac{\partial^2 u}{\partial y^2}\right) \quad and \quad \frac{\partial v}{\partial t} + u\frac{\partial v}{\partial x} + v\frac{\partial v}{\partial y} = -\frac{1}{\rho}\frac{\partial p}{\partial y} + \nu\left(\frac{\partial^2 v}{\partial x^2} + \frac{\partial^2 v}{\partial y^2}\right),
\end{equation}
\begin{equation}
    \frac{\partial u}{\partial x} + \frac{\partial v}{\partial y} = 0. \label{eq4}
\end{equation}

The continuity Eq. \ref{eq4} ensures mass conservation, which is critical for modeling the conservation of air and water masses in climate systems. The vorticity $\omega$, important for understanding cyclone formation \cite{rozanova2010typhoon}, and other climate phenomena \cite{farkane2023epinn}, is defined as:
\(    \omega = \nabla \times \mathbf{u} = \frac{\partial v}{\partial x} - \frac{\partial u}{\partial y}.\)
These equations collectively describe fluid behavior analogous to large-scale atmospheric and oceanic flows in climate systems \cite{SIMONNET2009187}. The spatial domain $X_{star}$ represents a discretized grid similar to those in climate models. The velocity field $U_{star}$ provides fluid motion data comparable to wind or ocean current measurements, while $p_{star}$ gives pressure distributions akin to atmospheric pressure patterns. The temporal evolution over 200 time steps allows for analyzing both short-term weather-like phenomena and longer-term climate-like behaviors, demonstrating the dataset's relevance to multiscale climate modeling challenges.

\subsection{Architecture}

\begin{figure*}
    \centering
    \includegraphics[width=\linewidth]{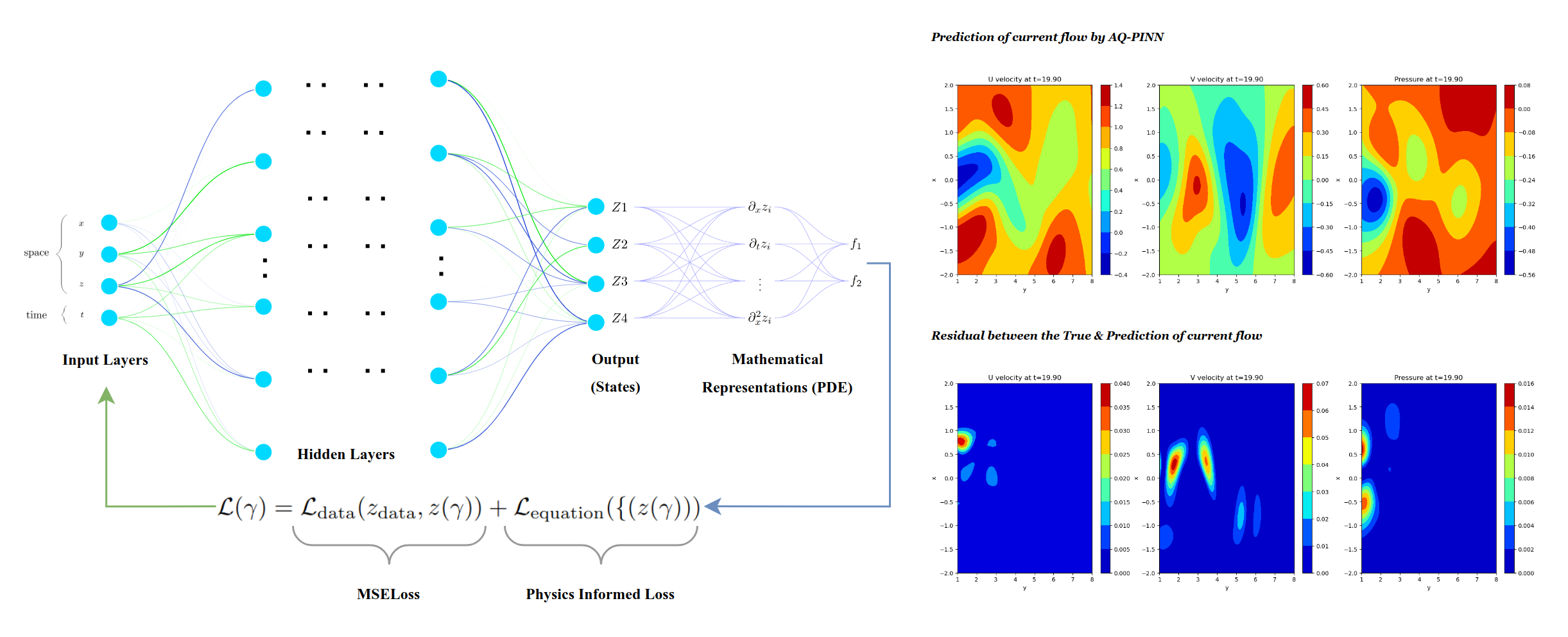}
    \caption{Residuals in the Navier-Stokes equations are computed precisely via \textbf{automated differentiation}. The loss function \(\mathcal{L}(\gamma)\) is differentiable for both the state variables \(z(\gamma)\) and the trainable parameters \(\gamma\) in AQ-PINNs. The predictions of AQ-PINNs can be understood from the diagrammatic representation in the right half.}
    \label{fig:AQ-PINN}
\end{figure*}

Physics-informed neural networks (PINNs) are a class of neural networks that integrate physical laws, typically expressed as partial differential equations (PDEs), directly into the learning process. They incorporate physical constraints to solve forward and inverse problems across various scientific and engineering disciplines. The key idea behind PINNs is to embed a system's governing equations into the neural network's loss function \cite{raissi2017physicsI,raissi2017physicsII,raissi2019physics}. 

The AQ-PINNs architecture consists of several key components, as shown in Fig. \ref{fig:AQ-PINN}. The input projection maps the input vector $\mathbf{x} \in \mathbb{R}^3$, representing spatiotemporal coordinates $(x, y, t)$, into a new space using a linear transformation. The core of AQ-PINNs is the Quantum Multi-head Self-Attention (QMSA) mechanism. This mechanism encodes classical data into quantum states, computes attention scores, and aggregates the results. Each input $x_i$ from the linear layer is encoded into a quantum state using a data loader operator $U^\dagger(x_i)$, where the unitary is represented as an angle embedding followed by circuits such as QMPS, QTTN, or QMERA for different benchmarks \cite{biamonte2019lectures,axioms13030187,10.1007/s10489-024-05337-w,dutta2024qadqn}.

For each input $x_i$, the key $K_i$, query $Q_i$, and value $V_{ij}$ are computed as:
\begin{equation}
K_i = \langle x_i|K^\dagger(\theta_K)Z_0K(\theta_K)|x_i\rangle, \quad Q_i = \langle x_i|Q^\dagger(\theta_Q)Z_0Q(\theta_Q)|xi\rangle, \nonumber
\end{equation}
\begin{equation}
    \quad V{ij} = \langle x_i|V^\dagger(\theta_V)Z_jV(\theta_V)|x_i\rangle,
\end{equation}

where $Z_0$ and $Z_j$ represent spin measurements of the qubit in the z-direction\cite{axioms13030187}. The attention matrix $A$ is then computed using the key and query vectors: \(A_{ij} = -(Q_i - K_j)^2.\)
The final output is obtained by applying the softmax function to the normalized attention matrix $\frac{A}{\sqrt{dh}}$ and multiplying it by the value matrix $V$ yielding to \( \textit{SoftMax}\left(\frac{A}{\sqrt{dh}}\right) \cdot V. \)
Two nonlinear transformations are applied using the $tanh$ activation function, allowing the model to learn nonlinear mappings while maintaining differentiability for gradient-based optimization followed by a final output projection which generates the predicted physical quantities $(\psi, \text{ssh}, u)$. The AQ-PINNs  model is trained using a composite loss function that combines data-driven and physics-driven components:

\begin{itemize}
    \item \textbf{Data Loss:} 
   \( \mathcal{L}_{\text{data}} = \mathbb{E}\left[(u - u_{\text{train}})^2 + (v - v_{\text{train}})^2 + (\text{ssh} - p_{\text{train}})^2\right] \),
    ensuring fidelity to the training data.

    \item \textbf{Physics-Informed Loss:} 
\( \mathcal{L}_{\text{phys}} = \mathbb{E}\left[(f_x - 0)^2 + (f_y - 0)^2 + (c - 0)^2\right] \),
    enforcing adherence to the Navier-Stokes equations, where $f_x$, $f_y$, and $c$ are the residuals computed using automatic differentiation \cite{farkane2023epinn}.
\end{itemize}

To optimize our training process, we utilize the L-BFGS method with a learning rate of $6.5E-1$, determined through the Super-Convergence technique. This approach involves gradually increasing the learning rate and identifying the value that results in the most rapid decrease in the loss function, ensuring swift and stable optimization \cite{smith2019super}. L-BFGS is a quasi-Newton method for solving second-order differential equations, such as the Navier-Stokes equations. By approximating the Hessian matrix, L-BFGS efficiently handles the intricate optimization landscape inherent in such equations, leading to quicker and more precise convergence \cite{raissi2017physicsII,cuomo2022scientific}.

\section{Results}

The proposed AQ-PINNs demonstrate a substantial reduction in model parameters while preserving and enhancing model performance in some instances. As illustrated in Table \ref{tab:model_performance}, the AQ-PINNs variants, utilizing distinct quantum tensor networks such as QMPS, QTTN, and QMERA, consistently achieved lower test loss values when compared to the classical attention-enhanced physics-informed neural networks (A-PINNs). Specifically, the AQ-PINNs models exhibit parameter reductions of up to 63.29\% (QMPS), 55.28\% (QTTN), and 51.51\% (QMERA), while simultaneously attaining test losses that are comparable to or better than those of the classical model. This significant reduction in model parameters is of particular importance in the domain of climate modeling, as it not only results in decreased computational demands but also contributes to a reduced carbon footprint.
\begin{table}[ht]
\footnotesize
    \centering
    \caption{Model performance and parameter reduction.}
    \begin{tabular}{|l|c|c|c|}
        \hline
        \textbf{Model} & \textbf{Test Loss Achieved} & \textbf{Decrease in Model Params} \\
        \hline
        Classical A-PINNs & 0.0631 & - \\
        AQ-PINNs (QMPS) & 0.0609 & 63.29\% \\
        AQ-PINNs (QTTN) & 0.0593 & 55.28\% \\
        \textbf{AQ-PINNs (QMERA)} & 0.0596 & \textbf{51.51\%} \\
        \hline
    \end{tabular}
    \label{tab:model_performance}
\end{table}

\section{Conclusion \& Future Work}

The paper introduces AQ-PINNs - Attention-enhanced Quantum Physics-Informed Neural Networks—offering a novel approach to carbon-efficient climate modeling by integrating quantum computing, physics-informed neural networks, and attention mechanisms targeting Jevons paradox. Our approach has shown promising results in improving model precision and computational efficiency. Moving forward, future work will focus on enhancing the interpretability of AQ-PINNs and address key challenges such as sub-grid process parameterization and extreme weather event prediction, aiming to refine and expand the model's capabilities in climate science. This includes developing advanced techniques for visualizing attention mechanisms, refining feature attribution methods, and creating more robust counterfactual explanations.
\section*{Acknowledgment}
This work was supported in part by the NYUAD Center for Quantum and Topological Systems (CQTS), funded by Tamkeen under the NYUAD Research Institute grant CG008.
\bibliographystyle{unsrt}
\bibliography{references}{}

\end{document}